\documentclass[fleqn,10pt]{wlscirep}
\usepackage[utf8]{inputenc}
\usepackage[T1]{fontenc}
\usepackage{physics}
\usepackage{IEEEtrantools}

\DeclareMathOperator*{\argmin}{arg\,min}

\title{Characterization and cancellation of power-line-induced motional-mode frequency noise in a trapped-ion system}

\author[1,2,3]{Jaehun You}
\author[1,2,3]{Jiyong Kang}
\author[1,2,3]{Kyunghye Kim}
\author[1,2,3]{Wonhyeong Choi}
\author[1,2,3,4,5,*]{Taehyun Kim}

\affil[1]{Department of Computer Science and Engineering, Seoul National University, Seoul, 08826, Republic of Korea}
\affil[2]{Automation and Systems Research Institute, Seoul National University, Seoul, 08826, Republic of Korea}
\affil[3]{NextQuantum, Seoul National University, Seoul, 08826, Republic of Korea}
\affil[4]{Institute of Computer Technology, Seoul National University, Seoul, 08826, Republic of Korea}
\affil[5]{Institute of Applied Physics, Seoul National University, Seoul, 08826, Republic of Korea}

\affil[*]{taehyun@snu.ac.kr}


\begin{abstract}
The stability of motional-mode frequency is essential for realizing high-fidelity quantum gates in trapped-ion quantum computing.
While broadband Gaussian noise has been extensively studied and mitigated using pulse shaping techniques, the impact of coherent periodic noise has remained largely unexplored.
Here we report a systematic investigation of 60-Hz power-line noise and its effect on the secular frequencies of a single ${}^{171}\mathrm{Yb}^{+}$ ion.
Using spin-echo Ramsey spectroscopy, we characterize the amplitude and phase of the resulting secular-frequency modulation and validate this characterization via passive phase correction of the Ramsey sequence.
Building on this, we implement a cancellation scheme by injecting a compensation tone into the set-point of a PI controller that stabilizes the trap RF drive amplitude.
A phasor-fitting procedure optimizes the amplitude and phase of the compensation signal, enabling near-complete suppression of the 60-Hz component.
With the cancellation applied, the coherence time of a radial motional mode is extended from approximately 10 ms to 35 ms, consistent with the limit set by motional heating.
Our results provide both a clear characterization of periodic motional-mode noise and a practical framework for its suppression in trapped-ion quantum computing platforms.
\end{abstract}
\begin{document}

\flushbottom
\maketitle
\thispagestyle{empty}

\section*{Introduction}

In trapped-ion quantum computers, the shared motional modes of the ions provide the quantum bus that mediates multi-qubit entangling gates~\cite{Cirac1995, Sorensen1999, Mintert2001, Leibfried2003_1}.
The long-term stability of these modes is therefore a central requirement for scalable, high-fidelity quantum logic operations~\cite{Leibfried2003_2, Bruzewicz2019, Katz2023}.
It is likewise essential for continuous-variable quantum computing and bosonic-encoding protocols that rely on precise phase-space control and phase accumulation of motional states~\cite{Fluhmann2019_1, Matsos2025}.

In many cases, dephasing in trapped-ion systems is modeled as stationary Gaussian noise and analyzed using the filter-function framework, where a control sequence acts as a frequency-domain filter on the noise power spectral density, enabling both noise spectroscopy and the systematic design of dephasing-robust operations~\cite{Soare2014, Green2015, Bermudez2017, Milne2020}.
More generally, this filter-function framework~\cite{Suter2016} has been widely applied across diverse platforms, including nuclear magnetic resonance~\cite{Alvarez2011}, superconducting circuits~\cite{Ithier2005, Bylander2011}, nitrogen-vacancy (NV) centers in diamond~\cite{Romach2015}, and semiconductor quantum dots~\cite{Connors2022}.

For motional modes in particular, frequency fluctuations are often characterized by a power spectral density~\cite{McCormick2019, Milne2021, Keller2021}.
This spectral characterization has enabled substantial progress in mitigating dephasing from stochastic Gaussian noise through noise-robust control techniques such as pulse shaping~\cite{Milne2020}.
However, trapped-ion experiments also face structured noise sources that are coherent and periodic, leading to deterministic modulation of the secular frequency with a well-defined phase relative to an external reference~\cite{Cai2021, deNeeve2022}.
Because the phase of such modulation is not captured by a stationary noise power spectral density alone, a common practical workaround is to phase-synchronize each experimental shot to the underlying periodic reference.

A prominent example of such structured noise is power-line noise at 50/60~Hz, which is encountered across diverse platforms, including superconducting circuits~\cite{Vepsalainen2022}, NV centers in diamond~\cite{Masuyama2021, Tayefeh2025}, and semiconductor quantum dots~\cite{Bharadwaj2025}.
In trapped-ion systems, power-line noise likewise constitutes a well-known structured noise source that can limit qubit coherence via power-line-synchronous magnetic-field fluctuations~\cite{Schindler2013}; accordingly, it is common practice to synchronize each experimental shot to a fixed phase of the line cycle (line triggering) to reduce its impact~\cite{Zhang2020, Wang2021, Kranzl2022, Gan2019}.
Building on this practice, several works have quantitatively characterized power-line-synchronous qubit-frequency modulation and demonstrated cancellation techniques to mitigate it~\cite{Merkel2019, Hu2023, Fluhmann2019_2}.
Motional-mode secular frequencies can exhibit analogous power-line-synchronous modulation through electrical pathways, for example via RF-drive amplitude pickup, and this has motivated the use of line-triggered operation in practice~\cite{Cai2021}.
Several works have further quantified this effect by characterizing the modulation~\cite{deNeeve2022}.
However, in contrast to the qubit case, practical methods for canceling line-synchronous secular-frequency modulation of motional modes appear to have received little attention to date.

In this work, we investigate the effect of 60-Hz power-line noise on the motional modes of a single trapped ${}^{171}\mathrm{Yb}^{+}$ ion, and we develop and demonstrate a systematic workflow for its characterization and cancellation.
By extracting both the amplitude and phase of the induced secular-frequency modulation, we implement a noise-cancellation scheme based on direct injection into the trap control system.
Our approach not only suppresses the dominant periodic component but also provides a practical template for identifying and compensating structured, coherent noise in trapped-ion systems.
To the best of our knowledge, this constitutes the first experimental demonstration of cancellation of periodic motional-mode noise in a trapped-ion system, highlighting both its practical significance and its relevance for future large-scale quantum processors.

\section*{Results}

\subsection*{Characterization of power-line noise}

\subsubsection*{Amplitude measurement using echo-based Ramsey spectroscopy}

We first benchmark motional coherence using Ramsey spectroscopy on the blue-sideband transition, starting from the qubit ground state and near the motional ground state, which is directly sensitive to secular-frequency fluctuations (see Supplementary Fig.~S1 online for experimental details).
The Ramsey signal as a function of the evolution time $\tau$ exhibits a Gaussian envelope over the first few milliseconds, $C(\tau) \propto \exp\!\left[-(\tau/T_g)^2\right]$, with a characteristic timescale $T_g$ associated with slow frequency fluctuations.
Such a non-exponential decay is characteristic of dephasing dominated by slowly varying frequency fluctuations, whereas fast Markovian (Lindblad-type) dephasing leads to an exponential envelope~\cite{Ithier2005, Suter2016}.
We therefore infer that Markovian dephasing is subdominant on the timescales probed here, and that the remaining decoherence is dominated by low-frequency contributions such as coherent periodic modulation and slow drift.
To suppress slow frequency wander on timescales of $\gtrsim 100$ ms while preserving sensitivity to coherent periodic modulation, we employ an echo-type Ramsey sequence~\cite{Hahn1950} (a single $\pi$ pulse at $\tau/2$) and multi-pulse Carr--Purcell (CP) sequences~\cite{Carr1954} with evenly spaced $\pi$ pulses.
Varying the number of $\pi$ pulses cross-validates the coherent-modulation signatures across different echo/CP orders while further suppressing slow drift.

In the presence of a sinusoidal modulation of the motional-mode secular frequency, the instantaneous mode frequency becomes $\omega(t) = \omega_0 + \delta\omega(t)$ with $\delta\omega(t) = A \cos(\omega_m t + \phi)$.
Here $A$ is the modulation amplitude (in rad/s, equivalently $A / 2\pi$ in Hz), $\omega_m = 2\pi f_m$ is the modulation angular frequency, and $\phi$ is the modulation phase; for power-line-synchronous modulation one typically has $f_m \approx 50/60~\mathrm{Hz}$.
Although a model including higher harmonics can better reproduce some finer features of the measured Ramsey traces, the present data do not uniquely determine the harmonic content.
We therefore use the single-tone model as the simplest robust description in the main text (see Supplementary Information, Sec.~5).
This time-dependent secular-frequency shift periodically modulates the phase accumulated during free evolution (i.e., the azimuthal phase of the Bloch vector on the equator), giving rise to revival features in the measured Ramsey signal as a function of the total evolution time $\tau$.
In our data, spin-echo Ramsey exhibits revivals at $1/30\,\mathrm{s}$ and its integer multiples, while multi-pulse CP sequences show revivals only at specific fractions such as $1/15\,\mathrm{s}$ and $1/10\,\mathrm{s}$.

We characterize the periodic modulation of a radial motional mode using CP sequences with $n = 1, 2, 3$ refocusing $\pi$ pulses ($n = 1$ corresponding to echo Ramsey).
The experiment is not synchronized to the 60-Hz power-line signal, so the measured Ramsey signals correspond to phase-averaged results.
In the sinusoidal-modulation model, the signal $C_n$ for each $n$ can be calculated analytically as (see Supplementary Information, Sec.~2 for the derivation)
\begin{IEEEeqnarray}{lCl}
    C_1(\tau) & = & J_0\!\left(\frac{A}{\omega_m} \cdot 4 \sin^2\!\left(\frac{\omega_m \tau}{4}\right)\!\right) \label{eq:C1} \\
    C_2(\tau) & = & J_0\!\left(\frac{A}{\omega_m} \cdot 8 \sin^2\!\left(\frac{\omega_m \tau}{8}\right) \sin\!\left(\frac{\omega_m \tau}{4}\right)\!\right) \label{eq:C2} \\
    C_3(\tau) & = & J_0\!\left(\frac{A}{\omega_m} \cdot 4 \sin^2\!\left(\frac{\omega_m \tau}{12} \right) \left(2 \cos\!\left(\frac{\omega_m \tau}{3}\right) - 1\right)\!\right) \label{eq:C3} ,
\end{IEEEeqnarray}
where $J_0$ is the Bessel function of the first kind of order zero, and $\tau$ is the total free-evolution time (sum of free-evolution intervals), excluding the finite durations of the $\pi/2$ and intermediate $\pi$ pulses.
Fitting these expressions to the data, including a motional-heating envelope, yields $A / 2\pi = 40\textrm{--}54~\mathrm{Hz}$ and a heating rate $\dot{\bar n} = 6\textrm{--}14$ for a radial mode with $\omega_0 / 2\pi \simeq 1.27~\mathrm{MHz}$.
While the analysis here focuses on coherent sinusoidal modulation, motional heating also contributes to the overall contrast decay.
The consideration of heating effects and their interplay with sinusoidal noise is described in detail in the Supplementary Information, Sec.~3, where we use the Quantum Toolbox in Python (QuTiP)~~\cite{Johansson2013}.

\begin{figure}[!htbp]
\centering
\includegraphics[width=\linewidth]{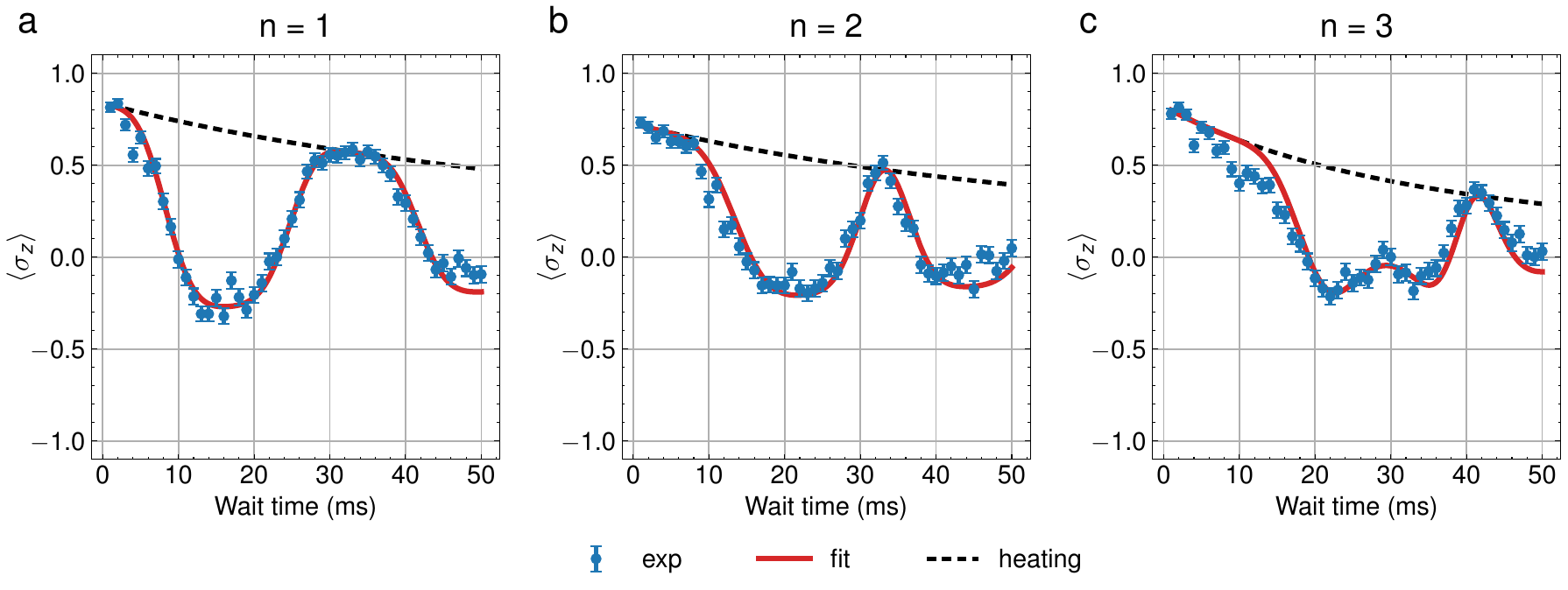}
\caption{
Amplitude characterization of sinusoidal secular-frequency modulation in a radial motional mode using CP sequences with (a) $n = 1$ (echo Ramsey), (b) $n = 2$, and (c) $n = 3$ refocusing pulses.
Blue circles show the measured Ramsey signal $\expval{\sigma_z}$ with error bars indicating shot noise ($N = 500$ shots per point).
Solid red curves show fits to the analytic sinusoidal-modulation model [Eqs.~(\ref{eq:C1})--(\ref{eq:C3})] including heating effects.
Best-fit parameters are (a) $A / 2\pi = 53.9(11)~\mathrm{Hz}$, $\dot{\bar n} = 6.4(8)~\mathrm{s}^{-1}$; (b) $A / 2\pi = 40.4(12)~\mathrm{Hz}$, $\dot{\bar n} = 7.1(15)~\mathrm{s}^{-1}$; (c) $A / 2\pi = 45.5(17)~\mathrm{Hz}$, $\dot{\bar n} = 13.6(26)~\mathrm{s}^{-1}$.
Dashed black curves show the independently simulated heating envelope alone; see Supplementary Information, Sec.~3 for simulation details.
Revivals occur at evolution times set by the 60-Hz line cycle: for echo Ramsey at integer multiples of $1/30\,\mathrm{s}$, and for multi-pulse CP at characteristic fractions (e.g., $1/15\,\mathrm{s}$ and $1/10\,\mathrm{s}$), as predicted by the corresponding $C_n(\tau)$ expressions.
}
\label{fig:char_ampl}
\end{figure}

Figure~\ref{fig:char_ampl} shows the measured Ramsey signal as a function of the evolution time $\tau$ for $n = 1, 2, 3$, together with fits of the model above to the data.
The revivals occur at evolution times set by the 60-Hz line cycle (e.g., for echo Ramsey at integer multiples of $1 / 30\,\textrm{s}$), consistent with a coherent line-synchronous modulation.

\subsubsection*{Phase measurement using delay-swept spin-echo Ramsey spectroscopy}

In this experiment, we operate in a phase-sensitive mode by synchronizing the start of each experimental shot to a fixed phase of the 60-Hz power-line signal.
We then introduce a programmable delay $t_d$ before the pulse sequence, so that the spin-echo Ramsey sequence is executed at a controllable phase of the line cycle (see Supplementary Fig.~S1 online for experimental details).
Modeling the resulting secular-frequency shift as $\delta\omega(t) = A \cos(\omega_m t + \phi_d)$, the delay deterministically maps to the modulation phase as $\phi_d = \phi_0 + \omega_m t_d$, where $\phi_0$ is the modulation phase at zero delay.
For each delay (i.e., modulation-phase) setting, we perform a spin-echo Ramsey measurement and fit the analytic expression
\begin{IEEEeqnarray}{c}
    C(\tau) = \cos\!\left(\varphi(\tau)\right), \quad
    \varphi(\tau) = \int_{0}^{\tau/2}\!A\,\cos(\omega_m t + \phi_d)\,\dd t\; - \!\int_{\tau/2}^{\tau}\!A\,\cos(\omega_m t + \phi_d)\,\dd t
\end{IEEEeqnarray}
to the measured Ramsey signal to extract the modulation phase $\phi_d$ for that delay.
Here $\varphi(\tau)$ is the phase accumulated during the echo sequence due to the sinusoidal secular-frequency shift.

\begin{figure}[!htbp]
\centering
\includegraphics[width=0.9\linewidth]{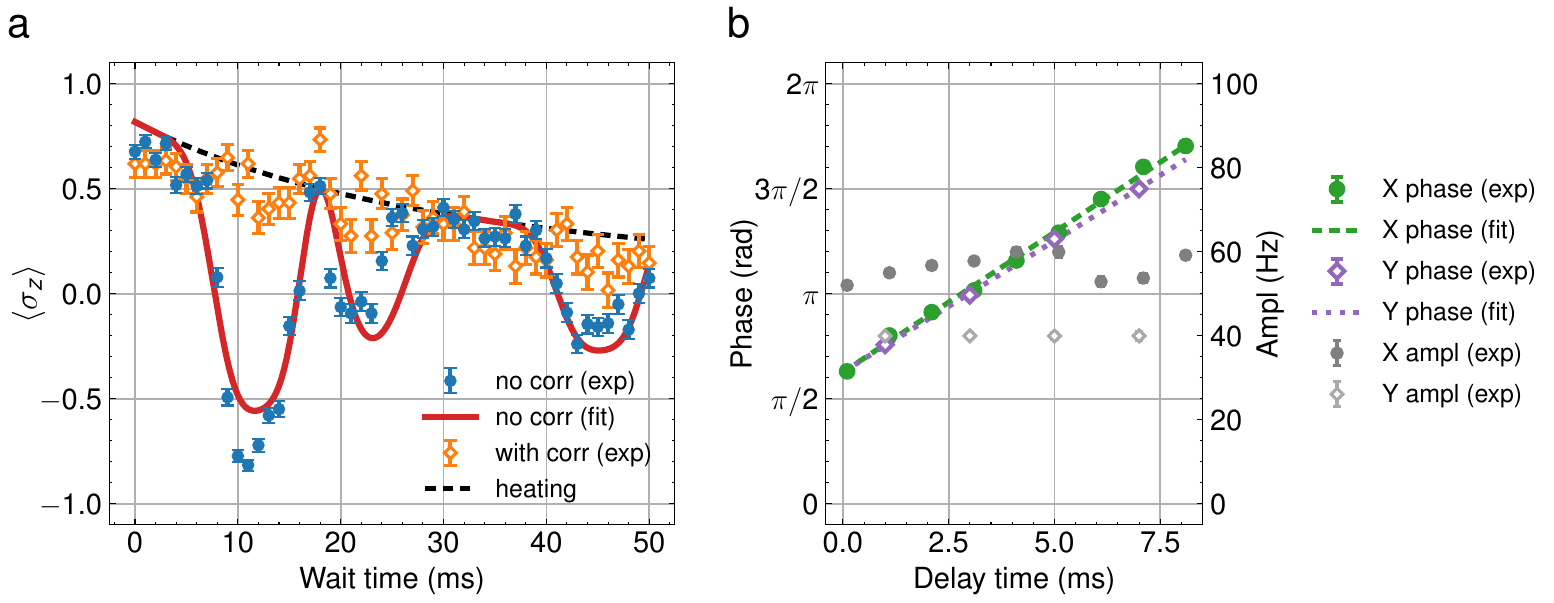}
\caption{
Phase characterization of sinusoidal secular-frequency modulation in radial motional modes using delay-swept spin-echo Ramsey measurements.
(a) Delay-swept spin-echo Ramsey measurement at a representative trigger delay of $2~\mathrm{ms}$.
The data are plotted in the same format as in Fig.~\ref{fig:char_ampl}: the experiment (blue circles with error bars indicating shot noise, $N = 500$ shots per point), a fit including sinusoidal noise and heating effects (solid red curve), and the heating effects alone (dashed black curve).
Best-fit parameters are $A / 2\pi = 56.8(7)~\mathrm{Hz}$, $\phi_d = 0.913(3)\pi$, and $\dot{\bar n} = 15.5(8)~\mathrm{s^{-1}}$.
Orange diamonds show the passive phase-correction verification with error bars indicating shot noise ($N = 150$ shots per point); see \textit{Verification via spin-echo Ramsey tracking}.
(b) Extracted modulation amplitude (dark-gray circle and bright-gray diamonds for the \textit{X} and \textit{Y} modes, respectively) and phase (green circle and purple diamonds for the \textit{X} and \textit{Y} modes, respectively) for the two radial modes as a function of the delay relative to the power-line trigger.
The fitted phase-evolution slopes (dashed green and dotted purple curves for the \textit{X} and \textit{Y} modes, respectively) are $2\pi \times 66.7(3)~\mathrm{s}^{-1}$ for the \textit{X} mode and $2\pi \times 62.5(9)~\mathrm{s}^{-1}$ for the \textit{Y} mode, both of which are close to the expected value of $2\pi \times 60~\mathrm{s}^{-1}$, confirming that the modulation is synchronous with the 60-Hz power-line signal.
}
\label{fig:char_phase}
\end{figure}

We find that the two radial modes (\textit{X} and \textit{Y}) exhibit different modulation amplitude $A$ while sharing essentially the same modulation phase $\phi_d$ as a function of the delay relative to the power-line trigger.
This in-phase behavior suggests that the 60-Hz modulation is coupled through the trap RF drive, producing a common-mode secular-frequency modulation in both modes.
Figure~\ref{fig:char_phase}a shows the delay-dependent spin-echo Ramsey fringes, while Fig.~\ref{fig:char_phase}b compares the extracted phases for the two modes.

\subsubsection*{Verification via spin-echo Ramsey tracking}

We verify the above characterization of the 60-Hz power-line noise by implementing a spin-echo Ramsey sequence in which the predicted phase accumulation during the free-evolution time is compensated.
Using the estimated power-line phase extracted from the delay-swept measurements in Fig.~\ref{fig:char_phase}, we calculate the expected phase accumulation during the free-evolution interval of the spin-echo Ramsey sequence.
This predicted phase is then applied to the phase of the second $\pi/2$ pulse, thereby passively compensating the phase shift induced by the 60-Hz modulation.
As shown by the orange diamond points in Fig.~\ref{fig:char_phase}a, applying this post-phase correction largely eliminates the dependence on the power-line noise, indicating that the extracted noise parameters capture the dominant contribution to the observed secular-frequency modulation.

\subsection*{Cancellation of 60-Hz power-line noise}

To cancel the 60-Hz power-line noise, we inject a sinusoidal compensation signal at the power-line frequency into the set point of the Proportional-Integral (PI) controller, as described in Methods.
When the compensation signal has the same amplitude but an opposite phase relative to the ambient 60-Hz noise, the resulting periodic secular-frequency modulation is canceled.

\begin{figure}[!htbp]
\centering
\includegraphics[width=0.7\linewidth]{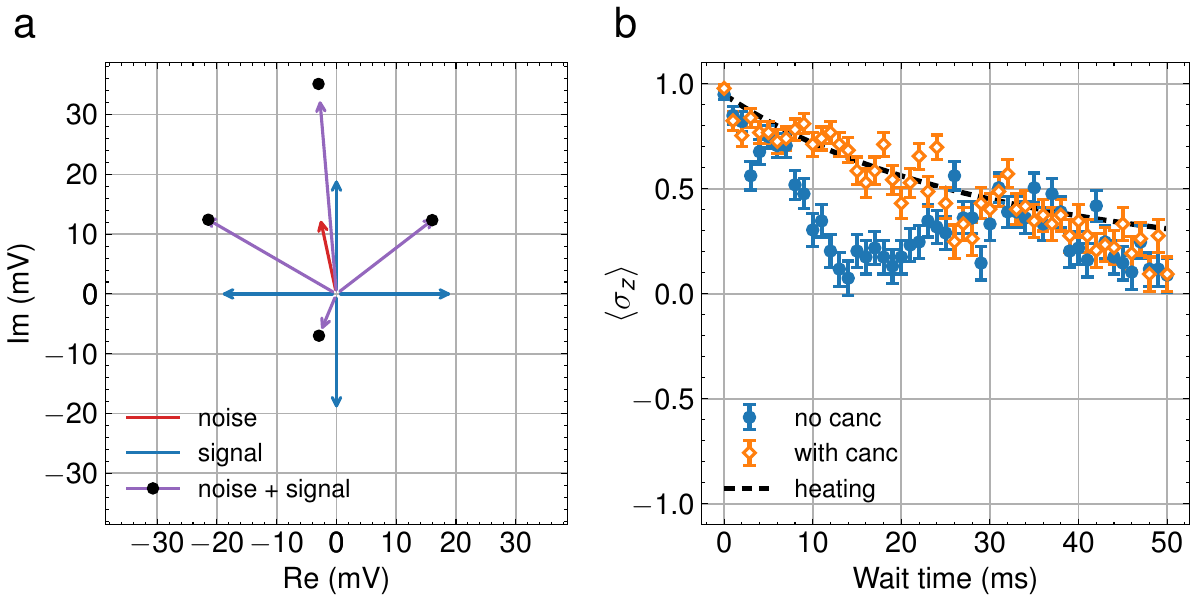}
\caption{
Cancellation of sinusoidal secular-frequency modulation using phasor-based optimization.
(a) Phasor diagram used for the least-squares optimization of the 60-Hz compensation signal.
Blue arrows denote the injected compensation phasors $V_i \hat{u}_i$ for four trial settings.
The red arrow indicates the extracted 60-Hz noise phasor $-V \hat{u}$.
The extracted noise phasor has amplitude $V = 14(3)~\mathrm{mV}$ and phase $\arg(-\hat{u}) = 102(10)^{\circ}$, and the scale factor is $r = 0.38(4)~\mathrm{mV/Hz}$.
Purple arrows show the resulting residual phasors $V_i \hat{u}_i - V \hat{u}$; black dots mark their endpoints, whose amplitudes are proportional to the measured residual modulation amplitude $A_i$ (after applying the scale factor $r$ defined in the text).
(b) Echo-Ramsey signal of a radial motional mode with and without cancellation.
Blue circles and orange diamonds show the data without and with cancellation applied, respectively, with error bars indicating shot noise ($N = 150$ shots per point) for both datasets; the dashed black curve shows the independently simulated heating envelope (using an average heating rate $\dot{\bar n} = 15~\mathrm{s}^{-1}$).
Cancellation suppresses the 60-Hz modulation and extends the coherence time from $\sim 10~\mathrm{ms}$ to $\sim 35~\mathrm{ms}$, close to the heating-limited value.
}
\label{fig:cancel}
\end{figure}

The amplitude of the 60-Hz component is first estimated from the characterization results above.
However, the phase of the compensation signal cannot be determined directly, because the injected 60-Hz tone is not phase-referenced to the experimental trigger; consequently, the relative phase between the injected tone and the ambient 60-Hz modulation is unknown a priori.
To overcome this, we perform a phasor-fitting procedure in which compensation signals with various amplitudes ($V_i$) and phases ($\hat{u}_i$) are sequentially injected into the controller.
Here $\hat{u}_i$ denotes a unit phasor ($|\hat{u}_i| = 1$) specifying the injected phase in the 60-Hz phasor plane.
For each trial signal, the residual 60-Hz modulation amplitude ($A_i$) is measured using spin-echo Ramsey sequences.
The relationship between the injected signal and the measured residual amplitude is then used to extract the optimal amplitude ($V$) and phase ($\hat{u}$) of the compensation signal by minimizing
\begin{IEEEeqnarray}{c}
    \argmin_{V, \hat{u}, r}\,\sum_i \left[\,\left|V \hat{u} - V_i\,\hat{u}_i\right| - r A_i\,\right]^2 .
\end{IEEEeqnarray}
Here $r$ is a proportionality constant that converts the measured modulation amplitude $A_i$ (in frequency units) to the corresponding phasor magnitude (in voltage units), accounting for the unknown transfer gain between the injected set-point modulation and the resulting secular-frequency modulation.
This optimization yields a vector representation of the 60-Hz noise phasor, allowing us to directly determine the opposite-phase compensation signal required for cancellation.
Figure~\ref{fig:cancel}a shows the phasor plot of the injected compensation signals and the corresponding residual amplitudes, from which the noise phasor is extracted.
Applying the resulting opposite-phase signal substantially suppresses the 60-Hz modulation in the spin-echo Ramsey measurements [Fig.~\ref{fig:cancel}b], demonstrating effective reduction of the dominant periodic noise component.
As a result, the coherence time of the radial motional mode increases from approximately 10~ms to 35~ms, close to the limit expected from the heating rate ($\dot{\bar n} = 15~\mathrm{s}^{-1}$).
Notably, once optimized, the same compensation settings have continued to suppress the 60-Hz component for at least one month, highlighting the suitability of the method for long-term operation.

\section*{Discussion}

The present cancellation scheme, which relies on a PI controller and an externally generated 60-Hz compensation signal, effectively suppresses the dominant power-line-synchronous noise in our system.
In our implementation, the compensation signal is optimized beforehand and then applied as a fixed waveform, which is sufficient because the dominant 60-Hz noise phasor remains stable over experimental timescales.
The implementation could be simplified and made more efficient.
Rather than using a dedicated signal generator, a 60-Hz reference could be derived directly from the power line using an electrically isolated sensing stage, followed by controllable phase shifting and amplitude scaling before injection into the controller set-point.
Alternatively, the power-line reference could be digitized with an ADC and processed in real time, and the resulting compensation waveform could be synthesized with a DAC for injection into the set-point.
Such an approach would reduce hardware complexity while preserving tunable amplitude and phase control for optimal cancellation.

Power-line-synchronous effects could in principle also arise from magnetic-field fluctuations via the second-order Zeeman shift of the qubit clock transition.
However, a Ramsey measurement on the clock transition does not show a comparably strong 60-Hz-synchronous signature, indicating that such magnetic-field-induced contributions are subdominant in our parameter regime (see Supplementary Information, Sec.~4).

The physical origin of the 60-Hz modulation is likely multifaceted.
In our setup, several essential devices---such as control electronics, RF amplifiers, and vacuum system components---must remain continuously connected to the power line.
Together with possible ground loops and electromagnetic pickup in the laboratory wiring, these connections can provide pathways for periodic interference to couple into the ion-trap electronics and modulate the effective confinement.

Even after cancellation of the 60-Hz component, residual fluctuations persist in the motional-mode secular frequencies.
In our measurements, the remaining noise is well described by a slow stochastic frequency drift with a characteristic scale on the order of tens of hertz (see Supplementary Information, Sec.~6).
Because this residual drift is not phase-referenced to the power line, it is not addressed by the present scheme.
Identifying the dominant mechanism and implementing complementary stabilization beyond cancellation of the power-line-synchronous component remain important directions for future work.

\section*{Methods}

\subsection*{Experimental setup}

We trap a single ${}^{171}\mathrm{Yb}^{+}$ ion in a blade-type linear Paul trap, as described in our previous work~\cite{Jeon2024, Jeon2025}.
The secular frequencies are $\omega_{X, Y, Z} \simeq 2\pi \times (910,\,1270,\,120)\,\mathrm{kHz}$; the axial mode at $120~\mathrm{kHz}$ is not used in this study.
The qubit is encoded in the hyperfine clock states $\ket{\downarrow} = \ket{S_{1/2}, F=0, m_F=0}$ and $\ket{\uparrow} = \ket{S_{1/2}, F=1, m_F=0}$.
State initialization and readout are performed using standard optical pumping and state-dependent fluorescence detection via the cycling transition, from which the population in $\ket{\uparrow}$ is inferred~\cite{Olmschenk2007}.
We prepare the relevant motional modes close to their ground states using resolved sideband cooling prior to each experiment.
The Raman beams used for spin-motion coupling are aligned such that the wavevector difference $\Delta\mathbf{k}$ is perpendicular to the trap axis, thereby coupling predominantly to the radial motional modes~\cite{Jeon2025}.

\subsection*{RF system and PI controller}

\begin{figure}[!htbp]
\centering
\includegraphics[width=0.95\linewidth]{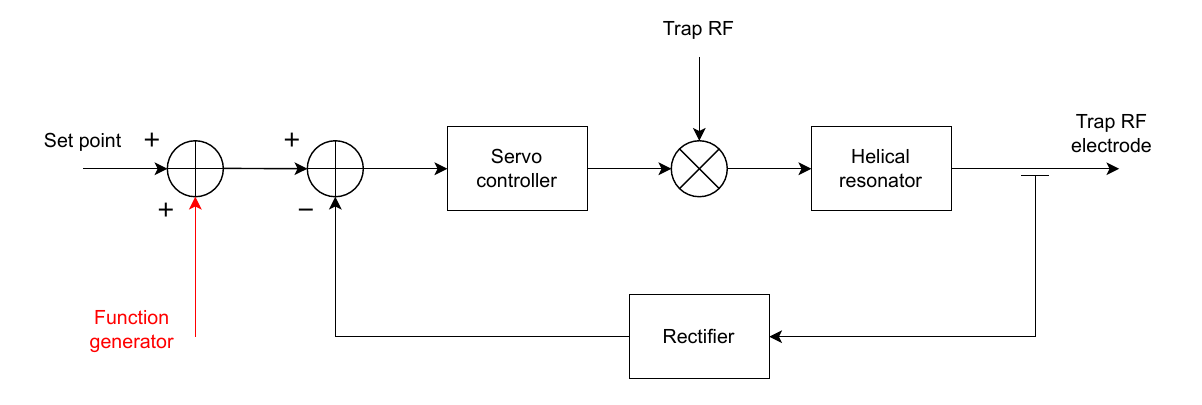}
\caption{
Schematic of the RF stabilization and compensation circuit.
The set-point is combined with an externally injected compensation tone, and the resulting reference is compared with a rectified monitor signal derived from the trap RF output.
The error signal drives a servo that controls the RF-drive amplitude via amplitude modulation of the RF source.
After filtering by the helical resonator, the RF is delivered to the trap electrodes, and a fraction of the RF is fed back for rectification and stabilization.
}
\label{fig:rf_circuit}
\end{figure}

The ion trap is driven by an RF source at approximately $17~\mathrm{MHz}$ and applied to the trap electrodes via a helical resonator.
The RF-drive amplitude is stabilized using a PI servo loop (New Focus LB1005 High-Speed Servo Controller), which adjusts the RF amplitude so that the measured amplitude tracks a set-point input; changes in the RF amplitude correspondingly shift the trap secular frequencies~\cite{Park2021}.
To enable cancellation of periodic noise, a periodic compensation signal is added to the set-point input of the PI controller, as illustrated schematically in Fig.~\ref{fig:rf_circuit}.

\subsection*{Cancellation implementation}

The compensation-signal amplitude is initially estimated relative to the PI controller set-point.
In our configuration, maintaining a secular frequency of approximately $1~\mathrm{MHz}$ requires a set-point voltage of about $1.43~\mathrm{V}$.
Assuming a locally linear response between the set-point and the secular frequency, an unwanted 60-Hz modulation with an effective magnitude of $\sim 50~\mathrm{Hz}$ corresponds to a set-point modulation on the order of $\sim 70~\mu\mathrm{V}$.
To generate a compensation signal at this level, we use a commercial signal generator in combination with fixed attenuation, adjusting the generator output amplitude such that a modulation on the order of tens of $\mu\mathrm{V}$ is applied at the set-point input.

In practice, the power-line frequency is not perfectly constant at $60~\mathrm{Hz}$ and can drift by up to $\sim \pm 1~\mathrm{Hz}$ over the course of an experiment.
Direct injection of a continuous 60-Hz tone would therefore lead to gradual phase slips between the compensation signal and the ambient 60-Hz modulation, reducing cancellation efficiency.
To address this, we employ the burst mode of the signal generator, following a previously reported approach~\cite{Hu2023}.
Using a trigger derived from the power line, the generator outputs a single-cycle sinusoidal burst at $62~\mathrm{Hz}$, i.e., one cycle at a slightly higher frequency than the nominal $60~\mathrm{Hz}$.
This ensures that the output waveform terminates before the next trigger event, allowing the phase to be re-referenced on every cycle and preventing accumulation of phase error.

\section*{Data availability}

The datasets generated during and/or analysed during the current study are available from the corresponding author on reasonable request.

\bibliography{references}

\section*{Author contributions statement}

J.Y. designed, carried out, and analyzed the experiments.
J.K. provided experimental guidance, particularly on the setup and control system.
J.Y., J.K., K.K., and W.C. contributed to the construction of the experimental apparatus.
T.K. supervised the project.
All authors reviewed and discussed the experimental results and contributed to the manuscript.

\section*{Funding}

This work was supported by the National Research Foundation of Korea (NRF) grant (No. RS-2024-00442855, No. RS-2024-00413957) and the Institute for Information \& Communications Technology Planning \& Evaluation (IITP) grant (No.RS-2022-II221040), all funded by the Korean government (MSIT).

\section*{Competing interests}

The authors declare no competing interests.

\end{document}


\maketitle
\vspace{0.5em}

\section{Experimental sequence}

\begin{figure}[!htbp]
\centering
\includegraphics[width=\linewidth]{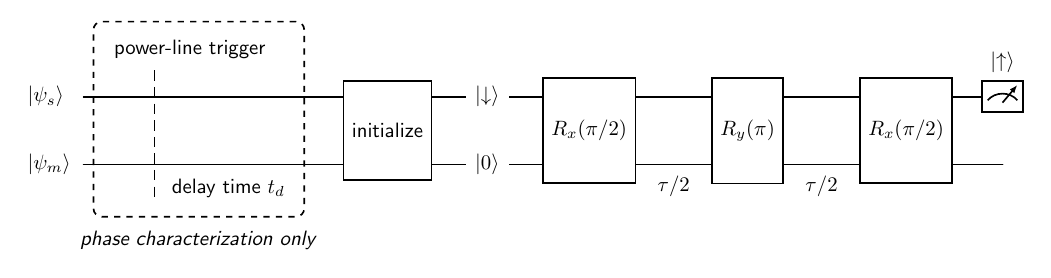}
\caption{
Experimental timing diagram for the full sequence (shown for the echo-Ramsey case with $n=1$).
Here $\ket{\psi_s}$ and $\ket{\psi_m}$ denote the spin-qubit and motional-mode states, respectively, and each shot is initialized to $\ket{\psi_s}\!\ket{\psi_m} = \ket{\downarrow}\!\ket{0}$ by qubit initialization and resolved sideband cooling.
The echo-Ramsey sequence is implemented on the blue-sideband transition between $\ket{\downarrow}\!\ket{0}$ and $\ket{\uparrow}\!\ket{1}$, with a single refocusing $\pi$ pulse at $\tau/2$.
The dashed box indicates the optional 60-Hz power-line trigger and programmable delay $t_d$ used only for phase characterization; it is omitted for amplitude characterization.
}
\label{fig:sequence}
\end{figure}

Figure~\ref{fig:sequence} illustrates the experimental sequence for the case of a single refocusing pulse ($n = 1$), i.e., the \emph{echo Ramsey} sequence.
After qubit initialization and resolved sideband cooling, the system is prepared in $\ket{\psi_s}\!\ket{\psi_m} = \ket{\downarrow}\!\ket{0}$, where the subscripts $s$ and $m$ denote the spin qubit and the motional mode, respectively.
A $\pi/2$ pulse on the blue-sideband transition then creates a coherent superposition between $\ket{\downarrow}\!\ket{0}$ and $\ket{\uparrow}\!\ket{1}$.
The system is then allowed to evolve for a total time $\tau$ with a refocusing $\pi$ pulse at $\tau/2$, and finally analyzed with a second $\pi/2$ pulse followed by state-dependent fluorescence detection.

For the \emph{amplitude characterization} measurements, we do not employ the trigger-delay block; each shot proceeds directly from initialization to the echo-Ramsey sequence.

For the \emph{phase characterization} measurements, as indicated by the dashed box in Fig.~\ref{fig:sequence}, we additionally synchronize the start of each experimental shot to a fixed phase of the 60-Hz power-line signal and insert a programmable delay time $t_d$ before the pulse sequence.

\newpage
\vspace*{-8mm}
\section{Echo-based Ramsey signal under sinusoidal secular-frequency modulation}

\noindent\textbf{Model and definition.}
We model the motional-mode secular-frequency shift as a sinusoidal modulation,
\begin{IEEEeqnarray*}{c}
    \delta\omega(t)
    = A \cos(\omega_m t + \phi),
\end{IEEEeqnarray*}
with modulation amplitude $A$ and angular frequency $\omega_m$.
The initial phase $\phi$ is assumed to be uniformly distributed on $[0, 2\pi)$.
We consider a Carr--Purcell (CP) sequence with $n$ refocusing $\pi$ pulses and define the toggling function $y_n(t) \in \{+1,-1\}$, which flips sign at each $\pi$ pulse~\cite{Hahn1950, Carr1954}.
The accumulated phase after total evolution time $\tau$ is
\begin{IEEEeqnarray*}{c}
    \varphi_n(\tau)
    = \int_{0}^{\tau} y_n(t)\,\delta\omega(t)\,\dd t.
\end{IEEEeqnarray*}
The phase-averaged Ramsey signal is given by
\begin{IEEEeqnarray*}{c}
    C_n(\tau)
    = \expval{\cos(\varphi_n(\tau))}_{\phi}.
\end{IEEEeqnarray*}
Using the identity $J_0(z) = \frac{1}{2\pi} \int_{0}^{2\pi} \cos(z\sin(\phi))\,\dd\phi$, we obtain
\begin{IEEEeqnarray*}{c}
    C_n(\tau)
    = J_0\!\left(\frac{A}{\omega_m}\,F_n(\omega_m \tau)\right),
\end{IEEEeqnarray*}
where $F_n$ is a dimensionless function determined by the pulse pattern.

\bigskip
\noindent\textbf{Ramsey ($n = 0$).}
With no refocusing pulse, $y_0(t) = +1$ on $[0,\tau]$.
\begin{IEEEeqnarray*}{rCl}
    \varphi_0(\tau)
    & = & A \int_{0}^{\tau}\!\cos(\omega_m t + \phi)\,\dd t
    \,=\, \frac{A}{\omega_m} \cdot 2 \sin\!\left(\frac{\omega_m}{2} \tau\right) \cos\!\left(\frac{\omega_m}{2} \tau + \phi\right) \\[3pt]
    C_0(\tau)
    & = & \frac{1}{2\pi} \int_{0}^{2\pi}\!\cos\!\left[\frac{A}{\omega_m} \cdot 2 \sin\!\left(\frac{\omega_m}{2} \tau\right) \cos\!\left(\frac{\omega_m}{2} \tau + \phi\right)\right] \dd\phi \\[3pt]
    & = & J_0\!\left(\frac{A}{\omega_m} \cdot 2 \sin\!\left(\frac{\omega_m}{2} \tau\right)\right),
\end{IEEEeqnarray*}
which yields $F_0(\vartheta) = 2 \sin(\vartheta/2)$.

\bigskip
\noindent\textbf{Echo-based Ramsey.}
With $n$ refocusing pulses, $y_n(t)$ alternates between $+1$ and $-1$ over the $n + 1$ free-evolution segments.
\begin{IEEEeqnarray*}{rCl}
    \varphi_n(\tau)
    & = & A \sum_{j = 0}^{n - 1} (-1)^j \left[\int_{\frac{j}{n} \tau}^{(\frac{j}{n} + \frac{1}{2n}) \tau}\!\cos(\omega_m t + \phi)\,\dd t
    \,- \int_{(\frac{j}{n} + \frac{1}{2n}) \tau}^{\frac{j + 1}{n} \tau}\!\cos(\omega_m t + \phi)\,\dd t\right] \\[3pt]
    & = & \frac{A}{\omega_m} \sum_{j = 0}^{n - 1} (-1)^j \cdot 2 \sin\!\left(\frac{\omega_m \tau}{4n}\right)
    \left[\cos\!\left(\left(\frac{j}{n} + \frac{1}{4n}\right)\omega_m \tau + \phi\right) - \cos\!\left(\left(\frac{j}{n} + \frac{3}{4n}\right)\omega_m \tau + \phi\right)\right] \\[3pt]
    & = & \frac{A}{\omega_m} \sum_{j = 0}^{n - 1} (-1)^j \cdot 4 \sin^2\!\left(\frac{\omega_m \tau}{4n}\right) \sin\!\left(\left(\frac{j}{n} + \frac{1}{2n}\right)\omega_m \tau + \phi\right) \\[7pt]
    \IEEEeqnarraymulticol{3}{l}{\left(\begin{array}{l}
    \text{Using the trigonometric sum identity,} \\[3pt]
    \sum_{j = 0}^{n - 1} (-1)^j \sin(a + jd) = \dfrac{\sin\!\left(\dfrac{n (d + \pi)}{2}\right)}{\cos\!\left(\dfrac{d}{2}\right)} \sin\!\left(a + \dfrac{(n - 1)(d + \pi)}{2}\right),
    \end{array}\right.} \\[7pt]
    & = & \frac{A}{\omega_m} \cdot 4 \sin^2\!\left(\frac{\omega_m \tau}{4n}\right)
    \frac{\sin\!\left(\dfrac{\omega_m \tau}{2} + \dfrac{n \pi}{2}\right)}{\cos\!\left(\dfrac{\omega_m \tau}{2n}\right)} \sin(\phi + \cdots) \\[20pt]
    C_n(\tau)
    & = & J_0\!\left(\frac{A}{\omega_m} \cdot 4 \sin^2\!\left(\frac{\omega_m \tau}{4n}\right)
    \frac{\sin\!\left(\dfrac{\omega_m \tau}{2} + \dfrac{n \pi}{2}\right)}{\cos\!\left(\dfrac{\omega_m \tau}{2n}\right)}\right),
\end{IEEEeqnarray*}
which yields $F_n(\vartheta) = 4 \sin^2(\vartheta/4n)
\dfrac{\sin(\vartheta/2 + n \pi/2)}{\cos(\vartheta/2n)}$.

\bigskip
For the sequences used in the experiment, the corresponding functions $F_n(\vartheta)$ are
\begin{IEEEeqnarray*}{rCl}
    F_1(\vartheta) & = & 4 \sin^2(\vartheta/4), \\[3pt]
    F_2(\vartheta) & = & 8 \sin^2(\vartheta/8) \sin(\vartheta/4), \\[3pt]
    F_3(\vartheta) & = & 4 \sin^2(\vartheta/12) (2 \cos(\vartheta/3) - 1).
\end{IEEEeqnarray*}

\newpage
\section{Heating effect on echo-based Ramsey signal}

We quantify how motional heating modifies the echo-based Ramsey signal by performing full density-matrix simulations.
The motional heating is modeled by a Lindblad term that increases the mean phonon number at a rate $\dot{\bar n} = 6~\mathrm{s}^{-1}$.
We numerically integrate the corresponding master equation using the Quantum Toolbox in Python (QuTiP)~\cite{Johansson2013}.

\begin{figure}[!htbp]
\centering
\includegraphics[width=\linewidth]{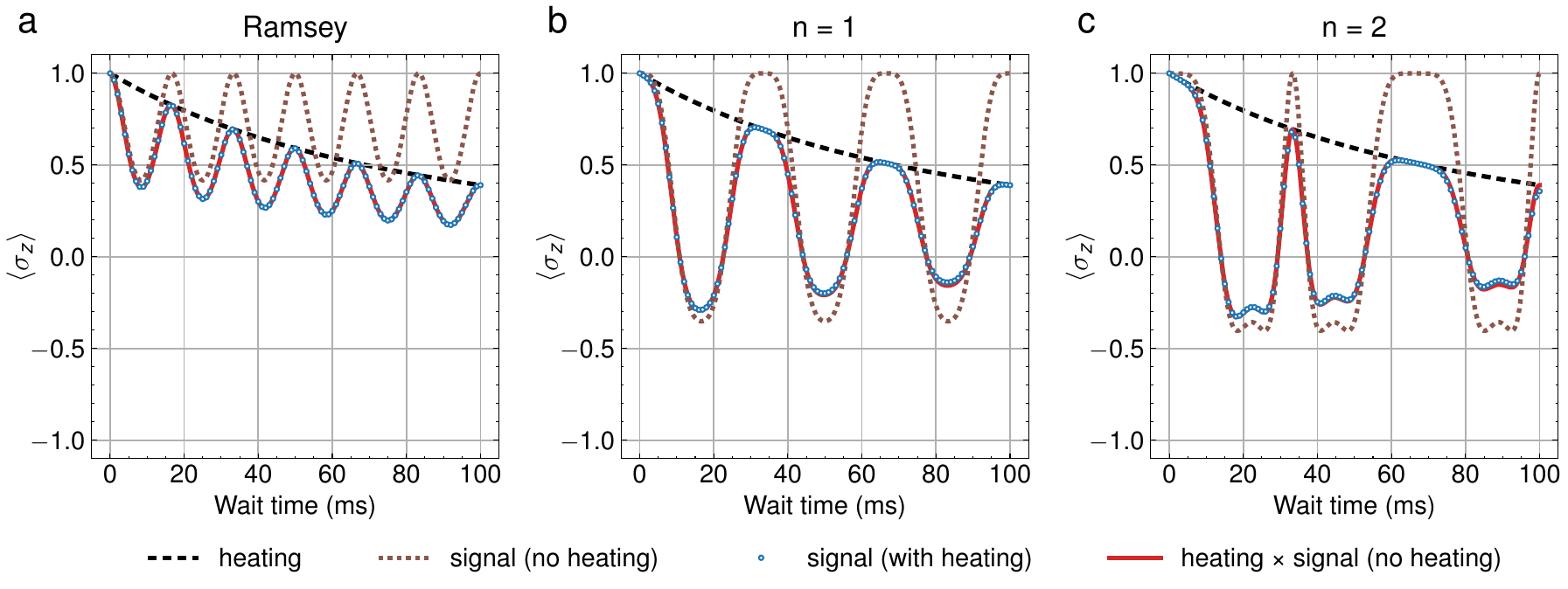}
\caption{
Effect of motional heating on echo-based Ramsey signals.
Dashed black curves: heating envelope $C_\mathrm{heat}(\tau)$ obtained from a Lindblad heating model with $\dot{\bar n} = 6~\mathrm{s}^{-1}$.
Dotted brown curves: simulated Ramsey signals without heating, $C_0(\tau)$, including the sinusoidal secular-frequency modulation.
Blue points: full simulations including both the sinusoidal secular-frequency modulation and motional heating, $C_\mathrm{tot}(\tau)$.
Solid red curves: product model $C_\mathrm{heat}(\tau)\,C_0(\tau)$, showing near-perfect agreement with the full simulation.
Panels (a--c) correspond to Ramsey, echo-based Ramsey ($n = 1$), and Carr--Purcell ($n = 2$), respectively.
}
\label{fig:heating_effect}
\end{figure}

To isolate the role of heating, we first compute a heating envelope $C_\mathrm{heat}(\tau)$ by simulating the same sequence with heating included but without sinusoidal secular-frequency modulation.
This envelope is shown as the dashed black curve in Fig.~\ref{fig:heating_effect} and captures the reduction of the Ramsey contrast due solely to heating.

Next, we compute the Ramsey signal in the absence of heating, $C_0(\tau)$ (dotted brown curve), which includes the sinusoidal secular-frequency modulation.
Finally, we perform the full simulation including both the secular-frequency modulation and motional heating to obtain $C_\mathrm{tot}(\tau)$ (blue points).

As demonstrated in Fig.~\ref{fig:heating_effect}, $C_\mathrm{tot}(\tau)$ is well approximated by the product of the heating envelope and the no-heating signal,
\begin{IEEEeqnarray*}{c}
    C_\mathrm{tot}(\tau)
    \approx C_\mathrm{heat}(\tau)\, C_0(\tau),
\end{IEEEeqnarray*}
shown as the solid red curve.
The near-perfect overlap indicates that, in our parameter regime, heating acts predominantly as a multiplicative amplitude decay and approximately factorizes from the sinusoidal secular-frequency modulation.
This factorization allows us to incorporate heating in fits using a separable envelope, while keeping the secular-frequency modulation model unchanged, enabling robust extraction of both the heating rate and modulation parameters.

\newpage
\section{Assessment of magnetic-field contributions via qubit clock-transition Ramsey}

\begin{figure}[!htbp]
\centering
\includegraphics[width=0.5\linewidth]{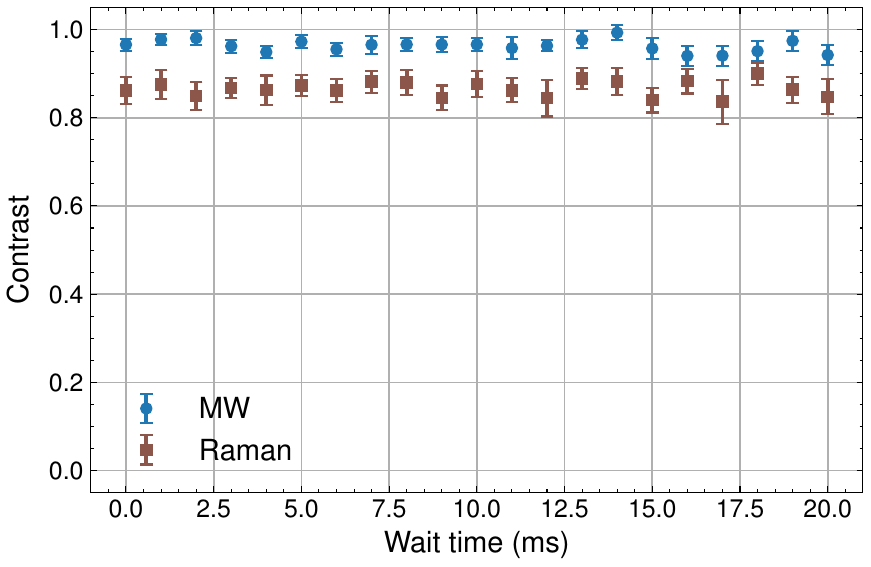}
\caption{
Ramsey contrast of the qubit clock transition as a function of the wait time, measured using microwave (blue circles) and Raman (brown squares) drives, with error bars indicating the fitting uncertainty.
The Raman data are acquired as a quick check under non-optimized calibration conditions and therefore show a lower overall contrast.
In both cases, the contrast remains nearly constant over the measured timescale and does not show a pronounced 60-Hz-synchronous feature.
}
\label{fig:qubit_ramsey}
\end{figure}

Power-line-synchronous modulation could in principle also arise from magnetic-field fluctuations.
Although the ${}^{171}\mathrm{Yb}^{+}$ clock transition, $\ket{S_{1/2}, F=0, m_F=0} \leftrightarrow \ket{S_{1/2}, F=1, m_F=0}$, is first-order insensitive to magnetic fields, a second-order Zeeman shift remains.
To assess whether this pathway could contribute appreciably to the line-synchronous signal observed in the motional-mode measurements, we perform Ramsey spectroscopy on the qubit clock transition using both microwave and Raman drives and extract the Ramsey contrast as a function of the wait time.
The resulting contrasts are shown in Fig.~\ref{fig:qubit_ramsey}.

Unlike the motional-mode Ramsey data in the main text, the qubit clock-transition Ramsey data do not show any pronounced oscillatory or revival feature associated with the 60-Hz line cycle for either microwave or Raman driving.
In both cases, the contrast remains nearly constant over the measured timescale, with no comparably strong line-synchronous signature.
These results therefore indicate that any power-line-synchronous magnetic-field-induced shift of the qubit transition is too small to account for the dominant 60-Hz response observed in the motional-mode Ramsey measurements.

This comparison supports the interpretation that the dominant effect discussed in this work arises primarily from secular-frequency modulation of the motional mode through the trap RF pathway, rather than from magnetic-field-induced shifts of the qubit transition.
While a small magnetic contribution cannot be completely excluded, the qubit Ramsey measurements indicate that such a contribution is subdominant in our parameter regime.

\newpage
\section{Higher-harmonic contributions to the line-synchronous modulation}

\bigskip
\noindent\textbf{Higher-harmonic model.}
Motivated by the possibility that the line-synchronous modulation may contain higher harmonics of the power-line frequency, we extend the single-tone model of Sec.~2 to
\begin{IEEEeqnarray*}{c}
    \delta\omega(t)
    = \sum_{k = 1}^{K} A_k \cos(k \omega_m t + \phi_k + \phi),
\end{IEEEeqnarray*}
where $k = 1$ corresponds to the 60-Hz component and $k > 1$ to its higher harmonics.
Here $\phi_k$ denotes the phase of the $k$-th harmonic relative to the fundamental component, with $\phi_1 = 0$, while $\phi$ represents the overall initial phase of the line-synchronous modulation for a given experimental shot.

The Ramsey signal under the $n$-pulse CP sequence can be evaluated in the same manner as in Sec.~2.
\begin{IEEEeqnarray*}{c}
    C_n(\tau) = J_0\!\left(\frac{4}{\omega_m} \sqrt{\left(\sum_{k = 1}^{K} \alpha_k \cos(\xi_k)\right)^2 +
    \left(\sum_{k = 1}^{K} \alpha_k \sin(\xi_k)\right)^2}\right),
\end{IEEEeqnarray*}
where $\alpha_k = \dfrac{A_k}{k} \sin^2\!\left(\dfrac{k \omega_m \tau}{4n}\right)
\dfrac{\sin\!\left(\dfrac{k \omega_m \tau}{2} + \dfrac{n \pi}{2}\right)}{\cos\!\left(\dfrac{k \omega_m \tau}{2n}\right)}$ and $\xi_k = \dfrac{k \omega_m \tau}{2} + \dfrac{(n - 1) \pi}{2} + \phi_k$.

\begin{figure}[!htbp]
\centering
\includegraphics[width=\linewidth]{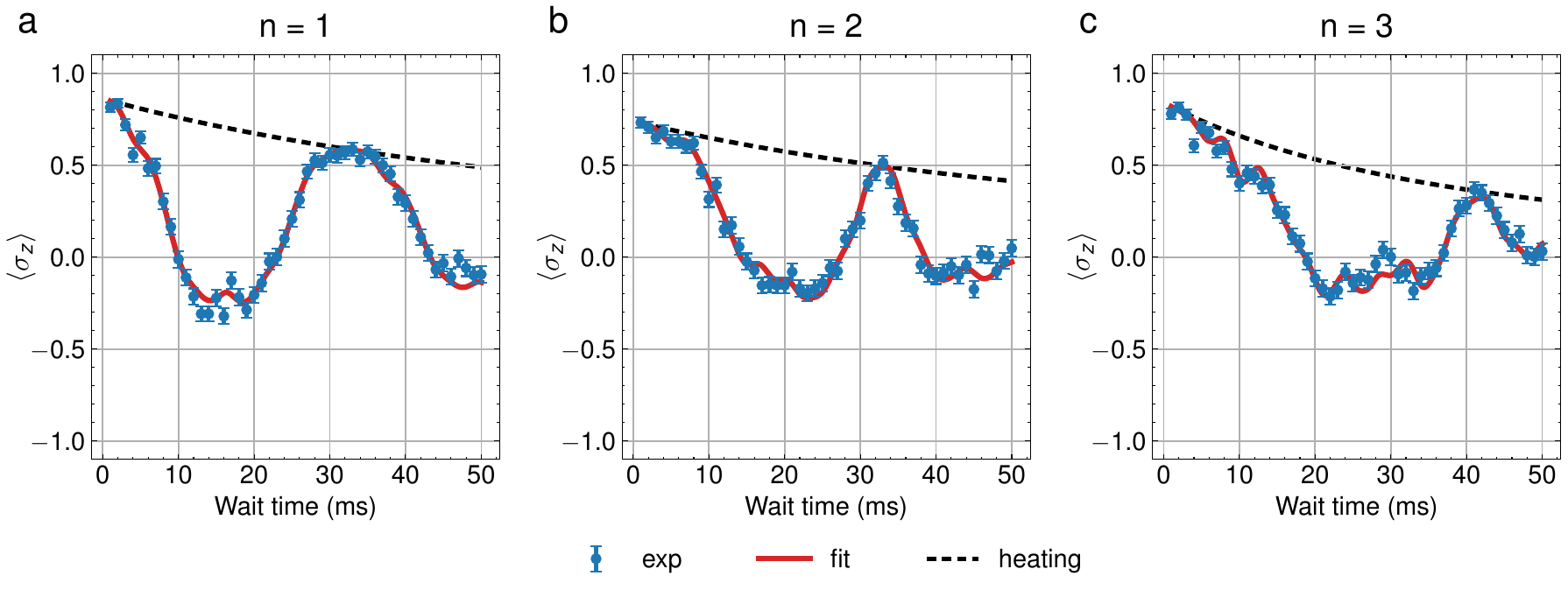}
\caption{
Fits of the amplitude-characterization data in a radial motional mode using the higher-harmonic model with harmonics up to $K = 5$, with (a) $n = 1$ (echo Ramsey), (b) $n = 2$, and (c) $n = 3$ refocusing pulses.
The data are plotted in the same format as in Fig.~1 of the main text: blue circles show the measured Ramsey signal \(\langle \sigma_z \rangle\) with error bars indicating shot noise (\(N = 500\) shots per point), and solid red curves show the lowest-loss fits obtained from 500 random initial parameter sets for each dataset.
Including higher harmonics better reproduces some finer features of the measured Ramsey traces for all three datasets.
}
\label{fig:char_ampl_harmonics}
\end{figure}

\bigskip
\noindent\textbf{Fits to the experimental data.}
Figure~\ref{fig:char_ampl_harmonics} shows representative fits of the $n = 1, 2, 3$ datasets using the model with harmonics up to $K = 5$.
For each dataset, we repeat the fit from 500 random initial parameter sets, and the result with the lowest fitting loss is shown.
The inclusion of higher harmonics better reproduces some finer features of the measured Ramsey traces for all three datasets, indicating that the line-synchronous modulation is not perfectly captured by a purely single-tone 60-Hz model.

Because the $K = 5$ model contains many free parameters, however, the detailed harmonic decomposition is not uniquely constrained by the present data.
Accordingly, the best-fit traces in Fig.~\ref{fig:char_ampl_harmonics} should be regarded primarily as representative examples showing that a higher-harmonic model can reproduce the observed signals, rather than as a unique extraction of the individual harmonic amplitudes and phases.
This point is examined further below by comparing a family of low-loss solutions.

\begin{figure}[!htbp]
\centering
\includegraphics[width=0.65\linewidth]{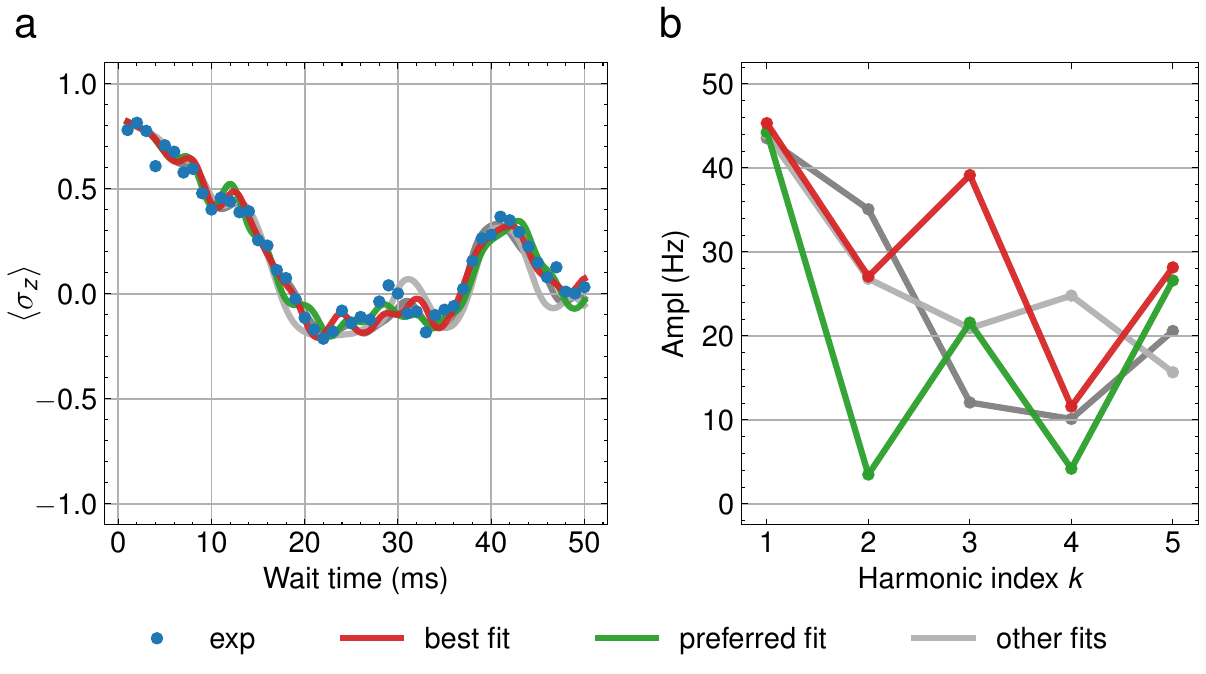}
\caption{
Illustration of the non-uniqueness of the higher-harmonic decomposition for the $n = 3$ dataset.
(a) Measured Ramsey signal (blue circles; error bars are omitted for clarity) together with the best-fit trace (solid red curve), a preferred low-loss fit (solid green curve), and two additional low-loss fits (solid gray curves), obtained from independent random initializations of the $K = 5$ model.
Here ``low-loss'' denotes solutions with fitting loss within a factor of 1.5 of the minimum value.
Although the fitted traces are nearly indistinguishable, the corresponding parameter sets differ substantially.
(b) Fitted amplitudes $A_k / 2\pi$ for the same family of low-loss solutions.
Only the amplitudes are shown here; the corresponding phases $\phi_k$ are omitted for clarity.
The 60-Hz component ($k = 1$) remains consistently prominent, whereas the higher-harmonic amplitudes show substantial variability.
The preferred fit is highlighted because its harmonic pattern is qualitatively more consistent with an independent oscilloscope measurement of the power line, which shows a dominant 60-Hz component and generally stronger odd-order than even-order harmonics.
}
\label{fig:harmonic_analysis}
\end{figure}

\bigskip
\noindent\textbf{Non-uniqueness of the harmonic decomposition.}
To visualize the non-uniqueness of the higher-harmonic fit, Fig.~\ref{fig:harmonic_analysis} compares a family of low-loss solutions for the $n = 3$ dataset.
Here, ``low-loss'' refers simply to solutions whose fitting loss lies within a factor of 1.5 of the minimum value obtained in the fit.
The best-fit solution, one preferred low-loss solution, and two additional low-loss solutions are selected from the 500 fits initialized from independent random starting points.
As shown in Fig.~\ref{fig:harmonic_analysis}a, these different parameter sets produce fitted traces that are nearly indistinguishable on the scale of the experimental data, despite corresponding to substantially different harmonic decompositions.

Figure~\ref{fig:harmonic_analysis}b shows the fitted amplitudes $A_k / 2\pi$ for the same set of low-loss solutions.
Here we display only the amplitudes and do not show the corresponding phases $\phi_k$, since the purpose of this plot is to illustrate the spread of the harmonic weights rather than to provide a complete visualization of the full parameter space.
The fitted amplitudes vary appreciably across the low-loss solutions, especially for the higher harmonics, whereas the fundamental component ($k = 1$, corresponding to 60 Hz) remains consistently prominent.

As an independent qualitative check, we also measure the harmonic content of the laboratory power line directly with an oscilloscope.
This measurement shows that the 60-Hz fundamental is the strongest component and that odd-order harmonics are generally stronger than even-order harmonics.
Among the low-loss solutions shown in Fig.~\ref{fig:harmonic_analysis}b, the preferred fit follows this qualitative trend more closely than the other admissible decompositions and is therefore likely to be closer to the actual power-line-noise spectrum in our setup. 
At the same time, these results indicate that, although the inclusion of higher harmonics can better reproduce some finer features of the measured Ramsey traces, the detailed harmonic decomposition is only weakly constrained by the present data alone.
We therefore use the single-tone 60-Hz model in the main text as the most parsimonious and robust description of the observed line-synchronous modulation.

\newpage
\section{Residual slow frequency drift}

\begin{figure}[!htbp]
\centering
\includegraphics[width=\linewidth]{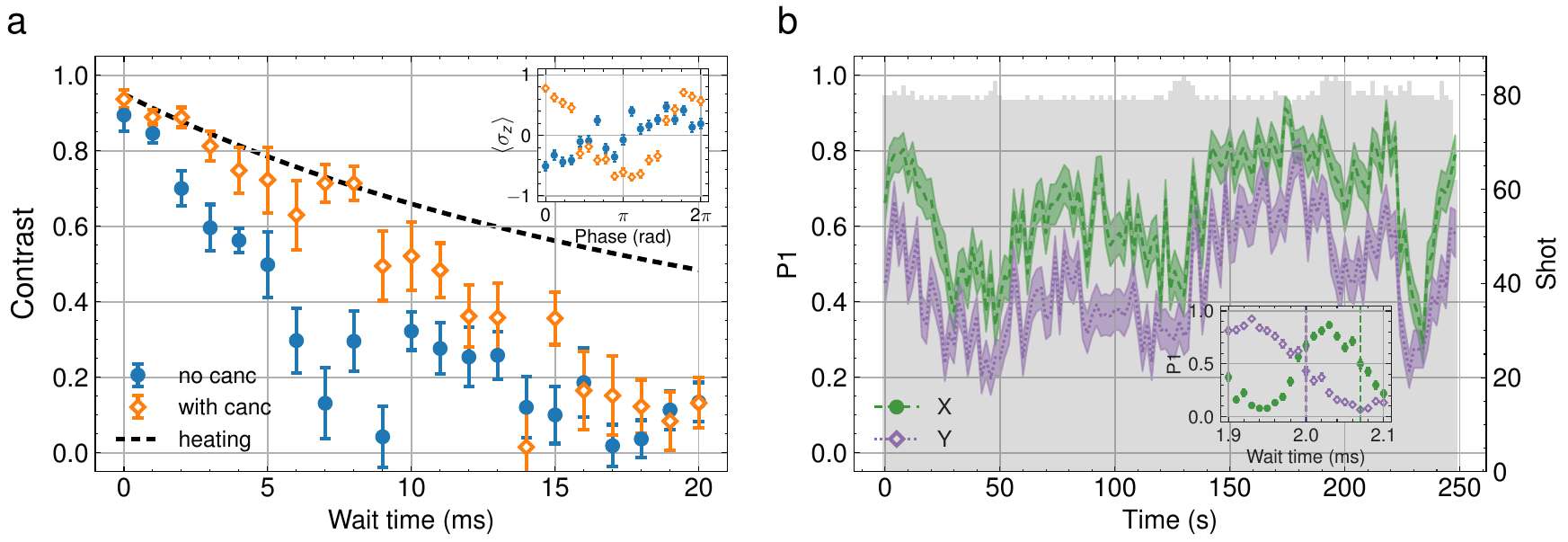}
\caption{
Residual dephasing from slow stochastic drift after cancellation of the 60-Hz component.
(a) Ramsey-contrast decay (no refocusing $\pi$ pulse) versus wait time for a radial motional mode, measured without cancellation (blue circles) and with cancellation (orange diamonds).
At each wait time, the contrast is extracted from a sinusoidal fit to a phase scan of the second $\pi/2$ pulse (inset).
The dashed black curve shows the independently simulated heating envelope.
(b) Time trace of single-shot Ramsey outcomes $P_1(t)$ at a fixed wait time near 2~ms for the two radial modes (dashed green and dotted purple curves for the \textit{X} and \textit{Y} modes, respectively), revealing slow shot-to-shot drifts.
Gray bars indicate the number of shots in each 2-s time bin.
Inset: Ramsey signal versus wait time near 2~ms; the wait time is chosen near $P_1 \approx 0.5$ to maximize sensitivity to small frequency shifts.
}
\label{fig:slow_drift}
\end{figure}

Although cancellation of the 60-Hz power-line noise extends the coherence time inferred from echo-Ramsey measurements to approximately 35~ms, this value does not reflect the practically usable timescale.
To probe the residual dephasing, we perform a Ramsey measurement without the refocusing $\pi$ pulse and quantify the Ramsey contrast as the wait time is varied.
Figure~\ref{fig:slow_drift}a shows the resulting contrast decay, together with the independently simulated heating envelope.
For each wait-time setting, we scan the phase of the second $\pi/2$ pulse to record a Ramsey fringe [inset of Fig.~\ref{fig:slow_drift}a] and extract the contrast from a sinusoidal fit to the phase-scan data.
Despite cancellation, the contrast exhibits a rapid decay beyond about 5~ms, indicating that additional dephasing processes remain.

To characterize this residual noise, we fix the Ramsey wait time to $\sim 2~\mathrm{ms}$ and repeat the measurement continuously over 250~s.
Figure~\ref{fig:slow_drift}b shows the resulting single-shot outcomes $P_1(t)$ for the two radial modes, where slow fluctuations reflect drifts of the motional-mode frequency.
As shown in the inset, we choose the wait time near $P_1 \approx 0.5$ (around 2~ms), where the Ramsey signal has maximal slope and is therefore most sensitive to small frequency shifts.
Assuming the residual noise is slow stochastic frequency drift, the observed Ramsey-signal fluctuations correspond to a standard deviation on the order of tens of hertz in the motional-mode secular frequency.
Such slow fluctuations impose a practical limitation on coherence, even when the 60-Hz component is suppressed.

\newpage
\bibliographystyle{naturemag-doi}
\bibliography{references}